\documentclass[twocolumn,aps,superscriptaddress,prb,longbibliography]{revtex4-1}
\setcitestyle{numbers,square}
\usepackage{amsmath,amssymb}
\usepackage{tabularx}
\usepackage{graphicx}% Include figure files \usepackage{dcolumn}% Align table columns on decimal point
\usepackage{bmpsize}
\usepackage{bm}% bold math
\usepackage{color}
\usepackage{amssymb}
\usepackage[version=3]{mhchem}
\usepackage{newtxmath}
\usepackage{physics}
\usepackage{hyperref}
\usepackage{mathtools}
\usepackage{comment}

\DeclareMathAlphabet{\mathpzc}{OT1}{pzc}{m}{it}

\begin{document}
%%%%%%%%%%%%%%%%%%%%%%%%%%%%%%%%%%%%%%%%%%%%%%%%%%%%%%%%%%%%%%%%%%%%%%%%%%%%%%%%%%%%%%%%%%%%%%%%%%%%%%%%%%%%%%%%%%%%%%%
\title{Landau levels and magneto-optics in 30$^\circ$ quasi-periodic twisted bilayer graphene}

\author{Masaru Hitomi}
\affiliation{Graduate School of Information Sciences, Tohoku University, Sendai 980-8579, Japan}
\author{Takuto Kawakami}
\affiliation{Center for Integrated Science and Humanities, Fukushima Medical University, Fukushima 960-1295, Japan}
\author{Mikito Koshino}
\affiliation{Institute for Solid State Physics, The University of Tokyo, Kashiwa, Chiba 277-8581, Japan}

\date{\today}
%%%%%%%%%%%%%%%%%%%%%%%%%%%%%%%%%%%%%%%%%%%%%%%%%%%%%%%%%%%%%%%%%%%%%%%%%%%%%%%%%%%%%%%%%%%%%%%%%%%%%%%%%%%%%%%%%%%%%%%

%%%%%%%%%%%%%%%%%%%%%%%%%%%%%%%%%%%%%%%%%%%%%%%%%%%%%%%%%%%%%%%%%%%%%%%%%%%%%%%%%%%%%%%%%%%%%%%%%%%%%%%%%%%%%%%%%%%%%%%
\begin{abstract}
We develop a theoretical framework for Landau levels in quasi-periodic twisted bilayer graphene at a $30^\circ$ twist angle, a system without translational symmetry but possessing 12-fold rotational symmetry. 
Using a quasi-band formalism, we incorporate the magnetic field through a conventional momentum substitution in the zero-field Hamiltonian. 
This approach provides a transparent physical interpretation by directly relating the Landau levels to the quasi-band structure, allowing them to be understood as quantized orbits of quasi-band pockets. 
By using this method, we reveal distinctive spectral features, including nearly flat bands with weak magnetic-field dependence and highly degenerate levels arising from twelve off-center pockets. 
The resulting Landau levels are classified by two quantum numbers: the Landau-level index and the angular momentum associated with the underlying quasicrystalline symmetry. 
We also compute the magneto-optical conductivity and show that optical transitions follow angular-momentum selection rules enforced by the 12-fold symmetry. 
Our approach provides a symmetry-based and computationally efficient framework for bulk quantum magneto-optics in quasicrystalline van der Waals systems, predicting spectroscopic signatures accessible in high-field infrared and THz experiments.
\end{abstract}
%%%%%%%%%%%%%%%%%%%%%%%%%%%%%%%%%%%%%%%%%%%%%%%%%%%%%%%%%%%%%%%%%%%%%%%%%%%%%%%%%%%%%%%%%%%%%%%%%%%%%%%%%%%%%%%%%%%%%%%
\maketitle
\section{Introduction}

In recent years, twisted bilayer systems of stacked two-dimensional (2D) materials with relative rotation angles have attracted considerable attention.  
The twisted structures exhibit physical properties markedly different from those of individual 2D layers, owing to moiré interference patterns in their atomic arrangements.  
A prominent example is low-angle twisted bilayer graphene (TBG), where the Dirac bands of pristine graphene are reconstructed into nearly flat bands
\cite{PhysRevLett.99.256802,PhysRevB.81.161405,bistritzer2011moire}, 
giving rise to superconductivity \cite{cao2018unconventional,yankowitz2019tuning,lu2019superconductors-b} and correlated insulating states \cite{cao2018correlated,lu2019superconductors-b}.  
Similar moiré-induced phenomena have also been reported in other van der Waals materials such as hBN \cite{yasuda2021stacking,woods2021charge}, transition metal dichalcogenides \cite{rivera2015observation,wu2019topological,seyler2019signatures}, and 2D magnets \cite{tong2018skyrmions,hejazi2020noncollinear}.

The moiré pattern and the associated electronic structure depend sensitively on the twist angle.  
At small angles (typically below $5^\circ$), the system exhibits a long-wavelength moiré periodicity, which allows treatment within the conventional Bloch formalism.  
As the angle increases, however, the interference length scale becomes comparable to the atomic spacing, and the system effectively loses translational periodicity.  
The resulting electronic states acquire a quasi-periodic character, reflecting the incommensurability of the stacked lattices.  

A particularly striking case is twisted bilayer graphene with a $30^\circ$ rotation ($30^\circ$ TBG).  
This structure realizes a quasicrystal with 12-fold rotational symmetry \cite{stampfli1986dodecagonal, Moon2019quasi,Sung2018dirac}.  
It has been experimentally synthesized 
\cite{Sung2018dirac,Wei2018quasi,yan2019scanning,suzuki2019ultrafast,pez202030tbg,liu2021synchronous,deng2020Interlayer,li2025Emergent}  
and theoretically investigated  
\cite{Moon2019quasi,yu2019dodecagonal,Koren2016sup,yu2020pressure,ha2021macroscopically,ghadimi2025Quasiperiodic,vidarte2024Quasicrystalline,yu2022Interlayer,park2019Emergent,crosse2021Quasicrystalline,spurrier2019Theory,li2020Quantum,yu2020electronic, vidarte2022high}.
In our previous work \cite{Moon2019quasi}, we demonstrated that a quasi-band description can be established in $30^\circ$ TBG, despite the absence of translational symmetry.  
The quasi-band structure exhibits 12-fold rotationally symmetric dispersions and flat-band features that are unattainable in periodic systems.

This raises a natural question: can one construct and analyze the Landau-level spectrum of such a quasi-band structure under an external magnetic field?  
In periodic crystals, weak-field spectra are obtained through the substitution $\bm{p}\to\bm{p}+e\bm{A}$ in the $\bm{k}\cdot\bm{p}$ Hamiltonian, where Landau levels can be interpreted as quantized orbits of Fermi pockets.  
For quasi-periodic systems, however, the absence of Bloch momentum precludes a straightforward application of this approach, and previous studies have instead relied on full Landau-level calculations using approximate methods such as finite-size models \cite{hatakeyama1987electronic,arai1987Electronic,hatakeyama1989fractal,tran2015topological}, commensurate approximants \cite{tran2015topological,fuchs2016hofstadter,fuchs2018Landau,yu2019dodecagonal,jeon2022Length,kozlov2023hofstadter}, and recursion techniques \cite{vidal2004quasiperiodic,vidarte2022high}.  
For $30^\circ$ twisted bilayer graphene, in particular, Landau levels have been theoretically investigated using these methods \cite{yu2019dodecagonal,yu2020electronic,vidarte2022high}, partially revealing the emergence of novel level structures \cite{yu2019dodecagonal,yu2020electronic}.  
However, their physical origin remains unclear due to the absence of an underlying band description, while the energy range associated with the characteristic 12-fold resonant quasi-band remains largely unexplored.

In this work, we develop an alternative and efficient framework to compute Landau levels in quasi-periodic systems and apply it to $30^\circ$ TBG.  
Our approach is based on the quasi-band formalism  
\cite{Moon2019quasi,koshino2015interlayer}:  
we introduce the minimal substitution $\bm{p}\to\bm{p}+e\bm{A}$ in the quasi-band Hamiltonian, in direct analogy with periodic systems.  
%This enables us to obtain the Landau-level spectrum without relying on commensurate approximants.  
Importantly, this method provides a transparent interpretation of the Landau-level structure in terms of the quasi-band structure, in which the levels are understood as quantized orbits of quasi-band pockets.  
In particular, the Landau levels can be classified by two quantum numbers: the 12-fold angular momentum associated with the underlying quasi-band and the Landau-level index.  

In the latter part of the paper, we compute the magneto-optical conductivity and uncover optical selection rules governed by the 12-fold angular momentum.  
Our results represent the first systematic derivation of Landau-level spectra in quasi-periodic systems, opening a pathway to their bulk quantum magneto-optics.  
We also note that this theoretical framework is not limited to $30^\circ$ TBG but is applicable to other quasiperiodic stacking systems.

The paper is organized as follows. 
In Sec.~\ref{sec_mod_30tbg}, we construct the effective Hamiltonian for $30^\circ$ TBG using the quasi-periodic approach of Ref.~\cite{Moon2019quasi}.  
In Sec.~\ref{sec_landau}, we introduce the Landau-level calculation method based on the quasi-band formalism and analyze the resulting spectrum of $30^\circ$ TBG.  
Section~\ref{sec_magopt} presents the calculation of the magneto-optical conductivity and discusses the associated optical selection rules.  
Finally, Sec.~\ref{sec_concl} provides a brief summary.

%%%%%%%%%%%%%%%%%%%%
\section{Quasi-band model for 30$^\circ$ TBG}
\label{sec_mod_30tbg}

\begin{figure}[h]
    \centering
    \includegraphics[width=0.9\hsize]{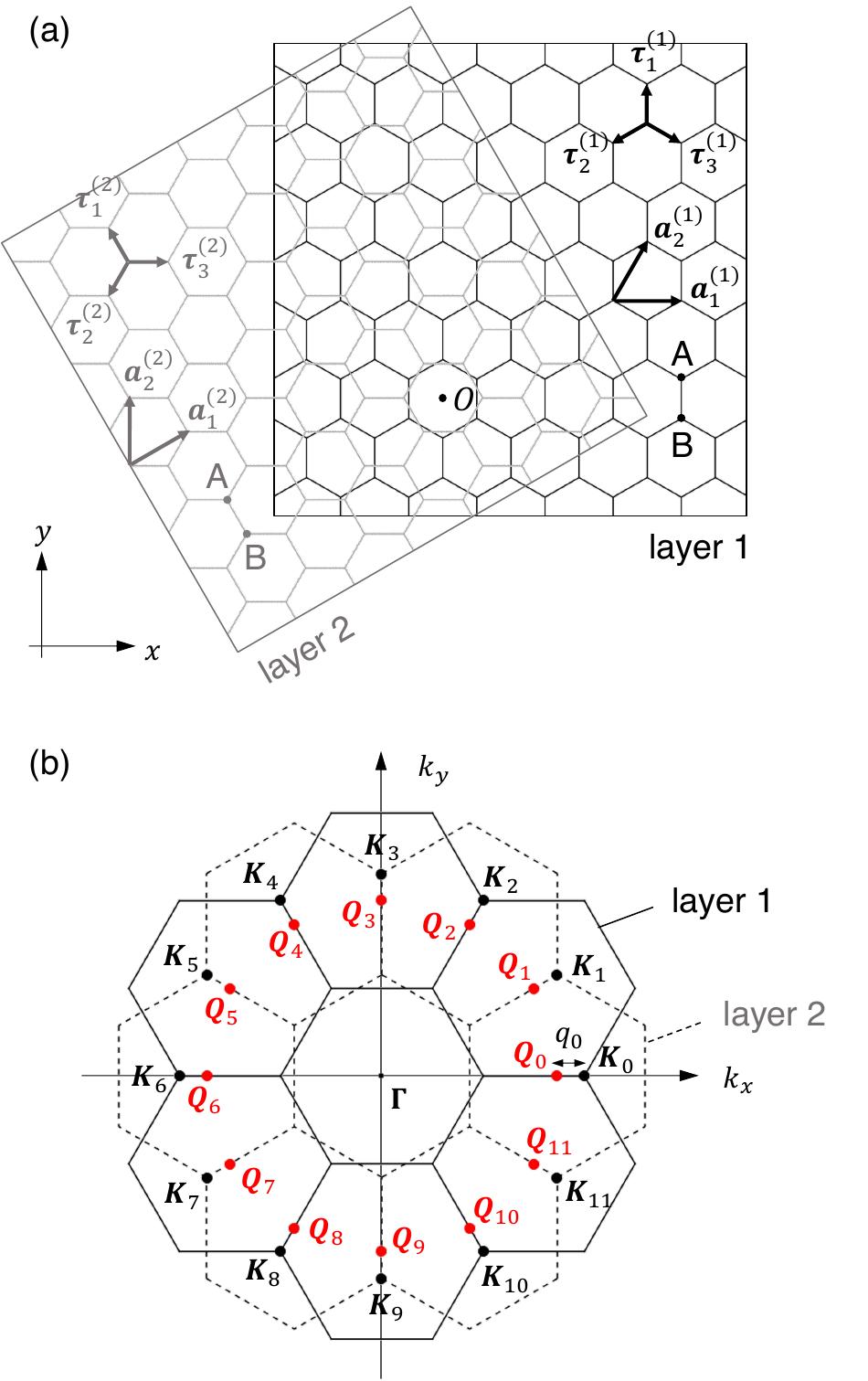}
    \caption{
    (a) Lattice structure of $30^\circ$ TBG.
    (b) Extended Brillouin zones of layer~1 (solid lines) and layer~2 (dashed lines).  
    Red dots mark the twelve symmetric points $\bm{Q}_n$ [Eq.~\eqref{eq_Qn}], which are folded onto the origin of the quasi-band structure in Fig.~\ref{fig_band_30tbg}.  
    Black dots $\bm{K}_n$ denote the nearest Brillouin-zone corners to $\bm{Q}_n$, corresponding to the Dirac points in Fig.~\ref{fig_band_30tbg}.
    }
    \label{fig_geom}
\end{figure}

\subsection{Tight-binding Hamiltonian}

We consider TBG with a twist angle of $\theta = 30^\circ$, as shown in Fig.~\ref{fig_geom}(a), 
where the graphene layers are modeled as flat, rigid honeycomb lattices of carbon atoms.
The rotation center is taken as the shared center of hexagon, which is set to be the origin.  
The primitive lattice vectors of layer 1 are defined by $\bm{a}^{(1)} _i = \bm{a}_i$
and those of layer 2 by  $\bm{a}^{(2)} _i =R(30^\circ) \bm{a}_i$ ($i=1,2$),
where $\bm{a}_1 = a(1,0)$ and $\bm{a}_2= a(1/2,\sqrt{3}/2)$  with $a \approx 0.246\,\mathrm{nm}$ are the lattice vectors of monolayer graphene and $R$ is the rotation matrix.
The corresponding primitive reciprocal lattice vectors of the individual layers are written as  $\bm{a}_i^{*(1)} =\bm{a}^*_i$ and $\bm{a}_i^{*(2)} =R(30^\circ)\bm{a}^*_i$,
where $\bm{a}^*_1 = (2\pi/a)(1,-1/\sqrt{3})$ and $\bm{a}^*_2=(2\pi/a)(0,2/\sqrt{3})$.
%The unit cell area of monolayer graphene is given by $S_0 = |\bm{a}_1\times \bm{a}_2|$.
We also define the vectors connecting neighboring carbon atoms $\bm{\tau}^{(1)} _i =\bm{\tau}_i$
and $\bm{\tau}^{(2)} _i =R(30^\circ)\bm{\tau}_i$ ($i=1,2,3$),
where
$\bm{\tau}_1 = (a/\sqrt{3})(0,1)$, 
$\bm{\tau}_2 =(a/\sqrt{3})(-\sqrt{3}/2,-1/2)$,
$\bm{\tau}_3 =(a/\sqrt{3})(\sqrt{3}/2,-1/2)$
[See, Fig.~\ref{fig_geom}(a)].

Each graphene layer contains two sublattices $A, B$ in its own unit cell.
The positions of sublattice $X(=A,B)$ on layer $l(=1,2)$ are given by 
\begin{align}
&\bm{R}^{(l)}_{X}=m_1\bm{a}^{(l)}_{1}+m_2\bm{a}^{(l)}_{2}+\bm{\tau}^{(l)}_{X}
\end{align}
Here $m_1$ and $m_2$ are integers, 
and  $\bm{\tau}^{(l)}_{X}$ is the relative sublattice position given by
$\bm{\tau}^{(l)}_{A}= -\bm{\tau}_1^{(l)} + z^{(l)}\bm{e}_z$,  $\bm{\tau}^{(l)}_{B}= \bm{\tau}_1^{(l)}+ z^{(l)}\bm{e}_z$.
Here $\bm{e}_z$ is the unit vector in the out-of-plane direction,
$z^{(1)}=0$ and $z^{(2)}=d_0$ are the heights of layer $l$,
where $d_0 = 0.334$ nm is the interlayer spacing.

We consider a single-orbital tight-binding model for the $p_z$ orbitals at carbon sites.
Let $|\bm{R}\rangle$ be a $p_z$ orbital at site $\bm{R}$.
We assume that the hopping energy between sites $\bm{R}$ and $\bm{R}'$ is expressed as
\begin{equation}
    \langle \bm{R}'| H | \bm{R} \rangle = -T(\bm{R}'-\bm{R}).
\end{equation}
We use the standard Slater-Koster form for $T(\bm{R})$,
\begin{align}
    \begin{split}
        -T(\bm{R}) 
        = V_{pp\pi} \left[1 - \left(\frac{\bm{R} \cdot \bm{e_z}}{R} \right)^2 \right]
        + V_{pp\sigma} \left(\frac{\bm{R} \cdot \bm{e_z}}{R} \right)^2, \\
        V_{pp\pi} 
        = V^{0}_{pp\pi} e^{-(R-a/\sqrt{3})/r_0},
        V_{pp\sigma} 
        = V^{0}_{pp\sigma} e^{-(R-d)/r_0},
    \end{split}
\end{align}
where $V^0_{pp\pi} \approx -2.7$ eV, $V^0_{pp\sigma} \approx -0.48$ eV, 
and $r_0 \approx 0.0453$ nm. \cite{trambly2010localization,moon2013optical} 

We define the Bloch bases as
\begin{align}
	& |\bm{k},X, l\rangle \equiv 
	\frac{1}{\sqrt{N}}\sum_{\bm{R} \in \bm{R}^{(l)}_{X}} e^{i\bm{k}\cdot\bm{R}}
	|\bm{R} \rangle
	\label{eq_bloch_base}
\end{align}
where $\bm{k}=(k_x,k_y)$ is a two-dimensional Bloch wave vector, 
and $N=S/S_0$ is the number of graphene's unit cells (the area $S_0 = |\bm{a}_1\times \bm{a}_2|$) per layer 
in the total system area $S$.
In this basis,
the intralayer matrix element of the Hamiltonian becomes
diagonal in $\bm{k}$, and it is written as
\begin{align}
   %h_{X'X}^{(l)}(\bm{k}) &\equiv  
   \langle\bm{k}, X',l| H |\bm{k}, X, l\rangle 
    &=  - \frac{1}{N} 
   \sum_{\bm{R} \in \bm{R}^{(l)}_{X}} 
   \sum_{\bm{R}' \in \bm{R}^{(l)}_{X'}} 
    T(\bm{R}' -\bm{R})\, e^{i\bm{k}\cdot(\bm{R}-\bm{R}')}.
    \label{eq_intra_1}
\end{align}
In this paper, we only consider the nearest neighbor hopping for the intralayer Hamiltonian, giving
\begin{align}
   &\langle\bm{k}, A,l| H |\bm{k}, B, l\rangle 
    =  - T_0 \sum_{i=1,2,3}  e^{-i\bm{k}\cdot\bm{\tau}^{(l)}_i},
    \nonumber\\
    &\langle\bm{k}, A,l| H |\bm{k}, A, l\rangle=\langle\bm{k}, B,l| H |\bm{k}, B, l\rangle=0,
    \label{eq_intra_approx}
\end{align}
where 
\begin{equation}
    -T_0 = V^0_{pp\pi} \approx -2.7\,{\rm eV}
\end{equation}
is the nearest neighbor hopping integral
in the monolayer graphene.

The interlayer matrix element is written as \cite{bistritzer2011moire,koshino2015interlayer}
\begin{align}
    & \langle \bm{k}^{(2)},X',2 |H| \bm{k}^{(1)},X,1 \rangle 
    \nonumber\\
    &=
     - \frac{1}{N}
    \sum_{\bm{R} \in \bm{R}^{(1)}_{X}} 
    \sum_{\bm{R}' \in \bm{R}^{(2)}_{X'}} 
    T(\bm{R}' -\bm{R})
    \, e^{i\bm{k}^{(1)}\cdot\bm{R}-i\bm{k}^{(2)}\cdot\bm{R}'}
    \nonumber\\
    &= -\sum_{\bm{G}^{(1)},\bm{G}^{(2)}} 
    t(\bm{k}^{(1)}+\bm{G}^{(1)})
    \, e^{-i\bm{G}^{(1)}\cdot\bm{\tau}^{(1)}_{X}
	+i\bm{G}^{(2)}\cdot\bm{\tau}^{(2)}_{X'}}
    \nonumber\\
    & \hspace{4cm} \times \delta_{ \bm{k}^{(1)}+\bm{G}^{(1)},\bm{k}^{(2)}+\bm{G}^{(2)}},
    \label{eq_interlayer_H}
\end{align}
where $\bm{G}^{(1)}=m_1\bm{a}^{*(1)}_1+m_2\bm{a}^{*(1)}_2$ and
$\bm{G}^{(2)}=m'_1\bm{a}^{*(2)}_1+m'_2\bm{a}^{*(2)}_2$
are the reciprocal lattice vectors of layer 1 and 2, respectively,
and $t(\bm{q})$ is a 2D Fourier transform of $T(\bm{R})$, or 
\begin{equation}
    t(\bm{q})
    = \frac{1}{S}
    \int{T(\bm{r}+ d_0 \bm{e}_z)
    e^{-i\bm{q} \cdot \bm{r}} d^2\bm{r}}.
    \label{eq_hop_tq}
\end{equation}
According to Eq. (\ref{eq_interlayer_H}), the interlayer coupling occurs between $\bm{k}^{(1)}$ of layer 1 and $\bm{k}^{(2)}$ of layer 2
satisfying the double Umklapp scattering condition 
$\bm{k}^{(1)} + \bm{G}^{(1)} = \bm{k}^{(2)} + \bm{G}^{(2)}$.
By multiple interlayer scatterings, an initial state at $\bm{k}$
can be coupled to a state at $\bm{k}+ \bm{G}^{(1)}+ \bm{G}^{(2)}$
for all $\bm{G}^{(1)}$ and $\bm{G}^{(2)}$.
Since $\bm{k}^{(l)}$ and $\bm{k}^{(l)} + \bm{G}^{(l)}$ stand for the same Bloch wavenumber of layer $l$,
the space of independent Bloch bases coupled by the interlayer Hamiltonian is written as \cite{Moon2019quasi}
\begin{align}
    & \{|\bm{k}^{(1)},X, 1\rangle \, ; \, \bm{k}^{(1)} = \bm{k} + \bm{G}^{(2)}, \forall\bm{G}^{(2)} \}
    \nonumber\\
    &\{|\bm{k}^{(2)},X, 2\rangle \, ; \, \bm{k}^{(2)} = \bm{k} + \bm{G}^{(1)}, \forall\bm{G}^{(1)} \},
    \label{eq_wavespace}
\end{align}
where $\bm{k}$ is an arbitrary wave number.

When the layer 1 and layer 2 are incommensurate, 
$\bm{G}^{(1)}$ and $\bm{G}^{(2)}$ never coincide,
and hence
$\bm{k} + \bm{G}^{(2)} (\bm{k} + \bm{G}^{(1)})$ represent all different Bloch wavenumbers of layer 1(2).
In typical twisted bilayer systems of our interest, however, the eigenstates do not infinitely spread out in the wave space,
but they are well described within a finite set of wave numbers of $\bm{G}^{(1)}$ and $\bm{G}^{(2)}$.\cite{Moon2019quasi}
The Hamiltonian can then be expressed as a finite matrix $\mathcal{H}(\bm{k})$ in the space of $\{|\bm{k}_n^{(l)},X, l\rangle\}$, where $l=1,2$, $X=A,B$, $\bm{k}_n^{(1)} = \bm{k} + \bm{G}_n^{(2)}$ and 
$\bm{k}_n^{(2)} = \bm{k} + \bm{G}_n^{(1)}$,
and $n$ is index to label $\bm{G}^{(l)}$ in the finite set.
The band dispersion of the system is obtained 
by plotting the eigenvalues of $\mathcal{H}(\bm{k})$ against $\bm{k}$.
This is referred to as a quasi-band in the following discussion.
The $\bm{k}$, denoted as the quasi-momentum, corresponds to the Bloch wavenumber in periodic systems.

\subsection{12-wave Hamiltonian}
\label{subsec_12fold}

The band structure of TBG at $\theta = 30^\circ$ can be well described by a 12-wave model \cite{Moon2019quasi}, which includes the Bloch bases at twelve $k$ points:
\begin{align}
& \{|\bm{k}_n^{(1)},X, 1\rangle\,;\, 
   \bm{k}_n^{(1)} =  \bm{k} + \bm{Q}_n \, (n=0,2,4,6,8,10)\},
    \nonumber\\
& \{|\bm{k}_n^{(2)},X, 2\rangle\,;\,
    \bm{k}_n^{(2)} = \bm{k} - \bm{Q}_n \, (n=1,3,5,7,9,11)\}.
    \label{eq_wavespace_12}
\end{align}
Here $X=A,B$ and $\bm{Q}_n$ is defined by
\begin{equation}
    \bm{Q}_n 
    = a^* \left[
    \cos{(n\pi/6)}, \sin{(n\pi/6)}
    \right],
    \label{eq_Qn}
\end{equation}
with $a^* = |\bm{a}^*_1| = |\bm{a}^*_2| = 4\pi/(\sqrt{3}a)$.
The positions of $\bm{Q}_n$ are illustrated in Fig.~\ref{fig_geom}(b).
The wave bases of Eq.~\eqref{eq_wavespace_12}
are a subset of Eq.~\eqref{eq_wavespace}, since $\bm{Q}_n (n={\rm even})$ belong to $\bm{G}^{(2)}$,
and  $\bm{Q}_n (n={\rm odd})$ to $\bm{G}^{(1)}$.
Using Eq.~\eqref{eq_intra_approx}, the intralayer matrix element is written as
\begin{align}
& \langle \bm{k}_{n}^{(1)},A, 1| H |\bm{k}_n^{(1)},B, 1\rangle
    =  - T_0 \sum_{i=1,2,3}  e^{-i(\bm{k}+\bm{Q}_n)\cdot\bm{\tau}^{(1)}_i},
    \nonumber\\
   & \hspace{5cm} (n=0,2,4,6,8,10)
     \nonumber\\
& \langle \bm{k}_{n}^{(2)},A, 2| H |\bm{k}_n^{(2)},B, 2\rangle
    =  - T_0 \sum_{i=1,2,3}  e^{-i(\bm{k}-\bm{Q}_n)\cdot\bm{\tau}^{(2)}_i},
     \nonumber\\
   & \hspace{5cm} (n=1,3,5,7,9,11)
    \label{eq_H_ring_intra}
\end{align}

Noting that the $\bm{k}_{n}^{(1)}$ and $\bm{k}_{n'}^{(2)}$ satisfy the coupling condition, $\bm{k}_{n}^{(1)} + \bm{G}^{(1)} = \bm{k}_{n'}^{(2)} + \bm{G}^{(2)}$ with $\bm{G}^{(1)} = - \bm{Q}_{n'}$ and $\bm{G}^{(2)} = \bm{Q}_{n}$,
the interlayer matrix element of Eq.~\eqref{eq_interlayer_H} is written as
\begin{align}
& \langle \bm{k}_{n'}^{(2)},X', 2| H |\bm{k}_n^{(1)},X, 1\rangle
\nonumber\\
& \qquad \qquad 
= - t(\bm{k} + \bm{Q}_n - \bm{Q}_{n'})
\, e^{i\bm{Q}_{n'}\cdot\bm{\tau}^{(1)}_{X}
	+i\bm{Q}_{n}\cdot\bm{\tau}^{(2)}_{X'}}.
\end{align}
Since $t(\bm{q})$ rapidly decays in large $\bm{q}$, the relevant coupling occurs when $\bm{k} + \bm{Q}_n - \bm{Q}_{n'}$ is small.
At $\bm{k}=0$, in particular, the strongest interlayer coupling occurs between $n$ and $n' = n\pm 1$ (mod 12) where $|\bm{Q}_n - \bm{Q}_{n'}|$ becomes the smallest, and the corresponding coupling amplitude is given by

\begin{align}
   w_0 \equiv t(\bm{Q}_n - \bm{Q}_{n+1}) = t(2a^* \sin{15^\circ}) \approx 0.157 \,{\rm eV}.
\end{align}
The effective Hamiltonian in the vicinity of $\bm{k}=0$ is obtained by only taking the coupling of $n$ and $n' = n\pm 1$  
and neglecting the $\bm{k}$ dependence as
$t(\bm{k} + \bm{Q}_n - \bm{Q}_{n+1}) \approx t_0$.

Consequently, the entire Hamiltonian can be expressed by a $24 \times 24$ matrix \cite{Moon2019quasi},
\begin{equation}
    \mathcal{H}(\bm{k})
    =
    \begin{pmatrix}
        H_0 & W^{\dagger} & & & & W \\
        W & H_1 & W^{\dagger} & & & \\
         & W & H_2 & W^{\dagger} & & \\
         & & \ddots & \ddots & \ddots & \\
         & & & W & H_{10} & W^{\dagger} \\
        W^{\dagger} & & & & W & H_{11}
    \end{pmatrix},
\label{eq_H_ring}
\end{equation}
\begin{align}
 &   H_n(\bm{k})
    =
    \begin{pmatrix}
       0 & h_n(\bm{k}) \\
       h_n^*(\bm{k}) & 0
    \end{pmatrix}, \quad
    W
    = -w_0
    \begin{pmatrix}
        \omega & 1 \\
        1 & \omega^* \\
    \end{pmatrix},
    \nonumber\\
&    h_n(\bm{k}) = h [R(-7n\pi/6)\bm{k}+\bm{Q}_0],
\label{eq_H_ring_block}
\end{align}
where  
\begin{equation}
    h(\bm{k}) = 
    - T_0 \sum_{i=1,2,3}  e^{-i\bm{k}\cdot\bm{\tau}_i},
    \label{eq_hk}
\end{equation}
and $\omega = e^{2\pi i/3}$.
The diagonal block $H_n$ represents
the monolayer's Hamiltonian at $\bm{k}_n^{(1)}$ for even $n$ (layer 1)
and $\bm{k}_n^{(2)} $ for odd $n$ (layer 2).
In each $2 \times 2$ matrix,
the sublattices are arranged in order of $(A, B)$ for $n \equiv 0, 3 \pmod 4$, and $(B, A)$ for $n \equiv 1, 2 \pmod 4$,
to obtain the rotationally symmetric form.
The expression of $H_n(\bm{k})$ in Eq.~\eqref{eq_H_ring_block} can be derived from Eq.~\eqref{eq_H_ring_intra} by noting that 
$\bm{Q}_n = \pm R(7n\pi/6) \bm{Q}_0$ for even and odd $n$, respectively, and also that the $7n\pi/6$ rotation of the set
of vectors $\{\bm{\tau}_1,\bm{\tau}_2,\bm{\tau}_3\}$
gives either of
$\pm\{\bm{\tau}^{(1)}_1,\bm{\tau}^{(1)}_2,\bm{\tau}^{(1)}_3\}$
or
$\pm\{\bm{\tau}^{(2)}_1,\bm{\tau}^{(2)}_2,\bm{\tau}^{(2)}_3\}$,
depending on $n$.

\begin{figure}[h]
    \includegraphics[width=1.\hsize]{fig_band_30tbg.pdf}
    \caption{
    (a) Quasi-band structure projected onto $k_y$ direction.
    (b) Enlarged view of the region enclosed by the dashed box in (a). 
    The green line represents the band structure at $k_x = 0$, and each value indicates the angular momentum of the band at $k = 0$. 
    The symbols $\pm$ denote degenerate states.
     (c) Three-dimensional plot of the off-centered 12-fold pockets in the valence band, corresponding to the vertical arrows in panel (a).}
    \label{fig_band_30tbg}
\end{figure}

The quasi-band structure of 30$^\circ$ TBG calculated by Eq.~\eqref{eq_H_ring} is illustrated in Fig.~\ref{fig_band_30tbg}(a).
The twelve Dirac cones originate from the monolayer's Hamiltonian $H_n$, where the origin $\bm{k}=0$ corresponds to 
$\bm{Q}_n (-\bm{Q}_n)$ of the Brillouin zone of graphene layer 1(2), shown in Fig.~\ref{fig_geom}(b).
The band touching point of $H_n$ for even (odd) $n$ occurs
when $\bm{k} + \bm{Q}_n (\bm{k} - \bm{Q}_n)$ coincides with 
a Brillouin corner point, $\bm{K}_n = (8\pi/3a)(\cos(n\pi/6),\sin(n\pi/6))$.
Consequently, in Fig.~\ref{fig_band_30tbg}(a), the Dirac cones are arranged on a circle of radius,
\begin{equation}
    q_0 
    = \frac{4\pi}{3a}(2-\sqrt{3}),
\end{equation}
which is the distance between $\bm{Q}_n$ and $\bm{K}_n$.
At $\bm{k} = \bm{0}$, 
the interlayer matrix element strongly couples the twelve degenerate states  
originally located at $\bm{Q}_n$ ($-\bm{Q}_n$) in the Brillouin zone of layer~1 (2) [Fig.~\ref{fig_geom}(b)].  
The resonant coupling of these 12-fold states gives rise to a characteristic dispersion with flat band edges
as shown in Fig.~\ref{fig_band_30tbg}(b) \cite{Moon2019quasi}. 
At the off-center points indicated by the vertical arrows, 
the Dirac cones centered at $\bm{K}_n$ and $\bm{K}_{n+2}$ \cite{Moon2019quasi} 
hybridize strongly, giving rise to the 12-pocket structure illustrated in Fig.~\ref{fig_band_30tbg}(c). 
These characteristic features are more pronounced in the valence band than in the conduction band.

For the Landau level calculation in the later section, we expand the Dirac Hamiltonian  $H^{(n)}$ in the 12-wave model [Eq.~\eqref{eq_H_ring_block}] around the Dirac point $\bm{K}_n$.
The expansion of the monolayer's Hamiltonian Eq.~\eqref{eq_hk} around $\bm{K}_0$ goes as
\begin{align}
    \begin{split}
        h(\bm{K}_0 + \delta\bm{k})
        &= - T_0
        \sum_{l=1}^{3}{
            e^{-i(\bm{K}_0 +\delta\bm{k}) \cdot \bm{\tau}_l}
        } \\
        &\approx -\hbar v
        (\delta k_x - i\delta k_y)
        -\frac{(\hbar v)^2}{6T_0}
        (\delta k_x + i\delta k_y)^2,
    \end{split}
    \label{eq_2nd_1}
\end{align}
where $\hbar v = (\sqrt{3}/2)a T_0$.
Accordingly, we have
\begin{align}
    & h_n(\bm{k})  \approx
    -\hbar v z_n - \frac{(\hbar v)^2}{6T_0}z_n^{*2},
           \nonumber\\
    & z_n =  -q_0 + e^{\frac{7}{6}i\pi n} (k_{x} - ik_{y}).
    \label{eq_2nd_2}
\end{align}
%In the following calculation, we adopt a reduced parameter $T_0=2.4$ eV 

\subsection{12-fold symmetry}
%\label{subsec_sym_30tbg}

The 12-fold symmetric nature of the band structure 
results from an improper rotational symmetry represented by $S_6=R(\pi/6)M_z$, which is a combination of 30$^\circ$ rotation around $z$ axis and the mirror reflection with respect to $xy$-plane.
This plays a central role in characterizing the Landau level structure and optical absorption spectrum.
For the Hamiltonian matrix of Eq.~\eqref{eq_H_ring},
the $S_6$ symmetry is expressed as
\begin{equation}
    S^{-1} \mathcal{H}(\bm{k}) S = \mathcal{H}[R({\pi}/6)\bm{k}],
\end{equation}
where
\begin{equation}
    S=
    \left(\begin{array}{cc}
     & I_{14\times14} \\
    I_{10\times10} & 
    \end{array}\right).
    \label{eq_s6mat}
\end{equation}
Here $S$ swaps the $n$-th 
$2 \times 2$ block to $n+7$-th block (mod 12) in a cyclic manner.
Since $\bm{k}=0$ is an invariant momentum under the $S_6$ operation, 
$\mathcal{H}(\bm{k}=0)$ commutes with $S$, and hence
it can be block-diagonalized with eigenvalues of $S$.
We define orthonormal eigenvectors
\begin{align}
    \bm{e}^A_m
    = \frac{1}{\sqrt{12}}\begin{pmatrix}
        e^{\frac{7 \cdot 0}{6}i\pi m} \\
        0 \\
        e^{\frac{7 \cdot 1}{6}i\pi m} \\
        0 \\
        e^{\frac{7 \cdot 2}{6}i\pi m} \\
        0 \\
        \vdots \\
        e^{\frac{7 \cdot 11}{6}i\pi m} \\
        0
    \end{pmatrix},
    \quad
    \bm{e}^B_m
    = \frac{1}{\sqrt{12}}\begin{pmatrix}
        0 \\
        e^{\frac{7 \cdot 0}{6}i\pi m} \\
        0 \\
        e^{\frac{7 \cdot 1}{6}i\pi m} \\
        0 \\
        e^{\frac{7 \cdot 2}{6}i\pi m} \\
        \vdots \\
        0 \\
        e^{\frac{7 \cdot 11}{6}i\pi m}.
    \end{pmatrix},
\label{eq_e_basis}
\end{align}
which satisfies $S \bm{e}_m^{X} = e^{-i m \pi/6} \bm{e}_m^{X}$.
Here $m=0, \pm 1, \pm 2,\cdots, \pm 5, 6$ correspond to the quantized angular momentum in 12-fold rotational symmetry.
We define a unitary matrix $U_m=(\bm{e}^A_{m}, \bm{e}^B_{m})$.

In this basis, the Hamiltonian Eq.~\eqref{eq_H_ring} with the quadratic approximation Eq.~\eqref{eq_2nd_2} can be unitary-transformed as
\begin{align}
     U_{m'}^{\dagger}\mathcal{H}(\bm{k})U_m 
    & = \tilde{H}_m \delta_{m',m}
   + \hbar v( k_+ \tilde{V}_1\,\delta_{m',m-1} 
   +  k_-\tilde{V}_1^\dagger\,\delta_{m',m+1})
   \nonumber\\
    & \quad
    + (\hbar v)^2(k_+^2\tilde{V}_2\,  \delta_{m',m-2} 
    + k_-^2\tilde{V}_2^\dagger\,  \delta_{m',m+2}),
    \label{eq_H_reduced_2nd}
\end{align}
where $k_\pm = k_x \pm i k_y$ and
\begin{align}
& \tilde{H}_m = \Bigl(\hbar v q_0 - \frac{(\hbar v q_0)^2}{6T_0}\Bigr)
 \begin{pmatrix}
   0&1 \\ 1&0
   \end{pmatrix}
-2w_0
   \begin{pmatrix}
   \cos \frac{7m-4}{6}\pi & \cos \frac{7m}{6}\pi
   \\ 
   \cos \frac{7m}{6}\pi &  \cos \frac{7m+4}{6}\pi
   \end{pmatrix},
   \nonumber\\
    & \tilde{V}_1 = 
   %\hbar v 
   \begin{pmatrix}
   0 &  \frac{\hbar vq_0}{3T_0} \\
   -1 & 0
   \end{pmatrix},
   \quad
   \tilde{V}_2 = 
  % -\frac{(\hbar v)^2}{6T_0}
   -\frac{1}{6T_0}\begin{pmatrix}
   0 & 1 \\
   0 & 0
   \end{pmatrix}.
   \label{eq_H_reduced_2nd-2}
\end{align}
We note that the matrix element from sector $m$ to $m+n\, (n=0,\pm1, \pm2)$ is accompanied by $(k_\mp)^n$, reflecting the conservation of the total angular momentum composed of the basis contribution $m$ from $\bm{e}^X_m$ and the orbital part in $\bm{r}$.

\section{Landau level spectrum}
\label{sec_landau} 

\subsection{Peierls substitution} 

In a tight-binding model, the effect of a magnetic field is incorporated by a Peierls substitution, which appends the phase factor $e^{i\phi_{\bm{R}',\bm{R}}}$, with
\begin{equation}
    \phi_{\bm{R}',\bm{R}} = -\frac{e}\hbar\int_{\bm{R}}^{\bm{R}'} \bm{A}(\bm{r}) \cdot d\bm{r},
    \label{eq_peierls}
\end{equation}
to the hopping integral $T(\bm{R}'-\bm{R})$ from site $\bm{R}$ to $\bm{R}'$. This approximation is valid when the vector potential $\bm{A}(\bm{r})$ varies smoothly on the atomic scale, i.e., for sufficiently weak magnetic fields.
As the electronic hopping occurs over an atomic distance, $\bm{A}(\bm{r})$ is nearly constant along the path in Eq.~\eqref{eq_peierls}, and we can approximate
\begin{equation}
    \phi_{\bm{R}',\bm{R}} \approx -\frac{e}\hbar\,\bm{A}\!\left(\frac{\bm{R}+\bm{R}'}{2}\right)\cdot(\bm{R}'-\bm{R}).
\end{equation}
For a conventional periodic system, the matrix element of a Bloch state is then modified as
\begin{align}
    \langle\bm{k}| H | \bm{k}\rangle
    &= \sum_{\bm{R},\bm{R}'} -T(\bm{R}'-\bm{R}) \,
       e^{i\phi_{\bm{R}',\bm{R}}} \,
       e^{i\bm{k}\cdot(\bm{R}-\bm{R}')}
    \nonumber\\
    &\approx \sum_{\bm{R},\bm{R}'} -T(\bm{R}'-\bm{R}) \,
    e^{\,i\left[\bm{k}+\tfrac{e}{\hbar}\bm{A}\!\left(\tfrac{\bm{R}+\bm{R}'}{2}\right)\right]
     \cdot(\bm{R}-\bm{R}')},
\end{align}
which coincides with the usual substitution rule in continuum systems, $\bm{k} \to \bm{k}+(e/\hbar)\bm{A}$.

A similar argument applies to the present quasiperiodic case.
The interlayer matrix element [Eq.~\eqref{eq_interlayer_H}] in the presence of a vector potential becomes
\begin{align}
    & \langle \bm{k}^{(2)},X',2 |H| \bm{k}^{(1)},X,1 \rangle 
    \nonumber\\
    &= - \frac{1}{N}
    \sum_{\bm{R} \in \bm{R}^{(1)}_{X}} 
    \sum_{\bm{R}' \in \bm{R}^{(2)}_{X'}} 
    T(\bm{R}' - \bm{R})
    e^{i\phi_{\bm{R}',\bm{R}}} 
    \, e^{i\bm{k}^{(1)}\cdot\bm{R}-i\bm{k}^{(2)}\cdot\bm{R}'}
    \nonumber\\
    &= - \frac{1}{N}
    \sum_{\bm{R} \in \bm{R}^{(1)}_{X}} 
    \sum_{\bm{R}' \in \bm{R}^{(2)}_{X'}} 
    T(\bm{R}' - \bm{R})
    \nonumber\\
    &\qquad\quad \times
        e^{\,i\left[\bm{k}+\tfrac{e}{\hbar}\bm{A}\!\left(\tfrac{\bm{R}+\bm{R}'}{2}\right)\right]\cdot(\bm{R}-\bm{R}')}
    \, e^{i\bm{G}^{(2)}\cdot\bm{R}-i\bm{G}^{(1)}\cdot\bm{R}'},
\end{align}
where we used $\bm{k}^{(1)} = \bm{k} + \bm{G}^{(2)}$ and
$\bm{k}^{(2)} = \bm{k} + \bm{G}^{(1)}$ [Eq.~\eqref{eq_wavespace}].
Thus, the vector potential enters as a shift of the quasi-momentum $\bm{k}$, as in a conventional Bloch system.

\subsection{12-wave Hamiltonian in magnetic field}

The 12-fold effective Hamiltonian in a magnetic field is obtained by applying $\bm{k} \to \bm{k}+(e/\hbar)\bm{A}$ in Eq.~\eqref{eq_H_reduced_2nd}. This is equivalent to replacing the momentum components $k_\pm = k_x \pm i k_y$ with the Landau-level ladder operators $a,a^\dagger$,  
\begin{align}
\label{eq_sbst}
     \hbar v k_+ \rightarrow \Delta_B a^{\dagger}, \,\,\,
     \hbar v k_- \rightarrow \Delta_B a,
\end{align}
where $\Delta_B = \sqrt{2\hbar v^2 e B}$.
After this substitution, we immediately notice that 
the Hamiltonian matrix preserves the total angular momentum $l = m + n$ (mod 12),
where $m$ is associated with $\bm{e}^{X}_{m}\, (X=A,B)$ and $n$ is the Landau level index related with the ladder operator $a$.
This motivates us to define
the following wave basis in magnetic field
\begin{equation}
    \ket{w^X_{l,n}} 
    \equiv \bm{e}^{X}_{l-n} \otimes \phi_n,
\label{eq_LLbasis}
\end{equation}
where $\phi_n (n=0,1,2\cdots)$ is the usual Landau level wavefunction, where the quantum number 
to label the degenerate states within the $n$-th level
is omitted.
Noting the operators $a^\dagger$, $a$ act on $\phi_n$, we have
\begin{align}
    a \ket{w^X_{l,n}}
    &= a \ket{\bm{e}^{X}_{l-n} \otimes \phi_n}
    = \sqrt{n} \ket{\bm{e}^{X}_{l-n} \otimes \phi_{n-1}},\nonumber\\
    a^{\dagger} \ket{w^X_{l,n}}
    &= a^{\dagger} \ket{\bm{e}^{X}_{l-n} \otimes \phi_n}
    = \sqrt{n+1} \ket{\bm{e}^{X}_{l-n} \otimes \phi_{n+1}}.
\end{align}

The Hamiltonian matrix elements in this basis are immediately obtained from Eqs.~\eqref{eq_H_reduced_2nd-2} and \eqref{eq_sbst} as
\begin{align}
     \bra{w^{X'}_{l',n'}} \mathcal{H} \ket{w^X_{l,n}} 
    & = \delta_{l',l} \Big[
    \tilde{H}_{l-n} \delta_{n',n}
    \nonumber\\
    & 
   + \Delta_B \sqrt{n+1}\, \tilde{V}_1\,\delta_{n',n+1} 
   +  \Delta_B \sqrt{n}\,\tilde{V}_1^\dagger\,\delta_{n',n-1}
   \nonumber\\
    &  
    +\Delta_B^2 \sqrt{(n+1)(n+2)}\, \tilde{V}_2\,\delta_{n',n+2} 
     \nonumber\\
    &  
   +  \Delta_B^2\sqrt{n(n-1)}\,\tilde{V}_2^\dagger\,\delta_{n',n-2}
   \Big],
    \label{eq_H_LL_basis}
\end{align}
where $\tilde{H}_{l-n}, \tilde{V}_1$, and $\tilde{V}_2$ are $2\times2$ matrices defined in Eq.~\eqref{eq_H_reduced_2nd-2}, and $X'$ and $X$ denote the indices for the row and column of these matrices, respectively.
Since the Hamiltonian preserves the total angular momentum $l$, the Hamiltonian matrix $\mathcal{H}$ is block-diagonalized into sub-Hamiltonians corresponding to $l = 0, 1, 2, \ldots, 11$.
In the following, we compute the eigenstates independently for each $l$, 
with a cutoff introduced for the Landau level index, 
$n \leq n_{\mathrm{max}} = 4000$.
The overall Landau-level spectrum of the system is obtained by 
collecting the eigenvalues from all sub-Hamiltonians 
with $l$.

\subsection{Landau levels of 30$^\circ$ TBG}
\label{subsec_sp_LL}

\begin{figure*}
    \centering
    \includegraphics[width = 1.\hsize]{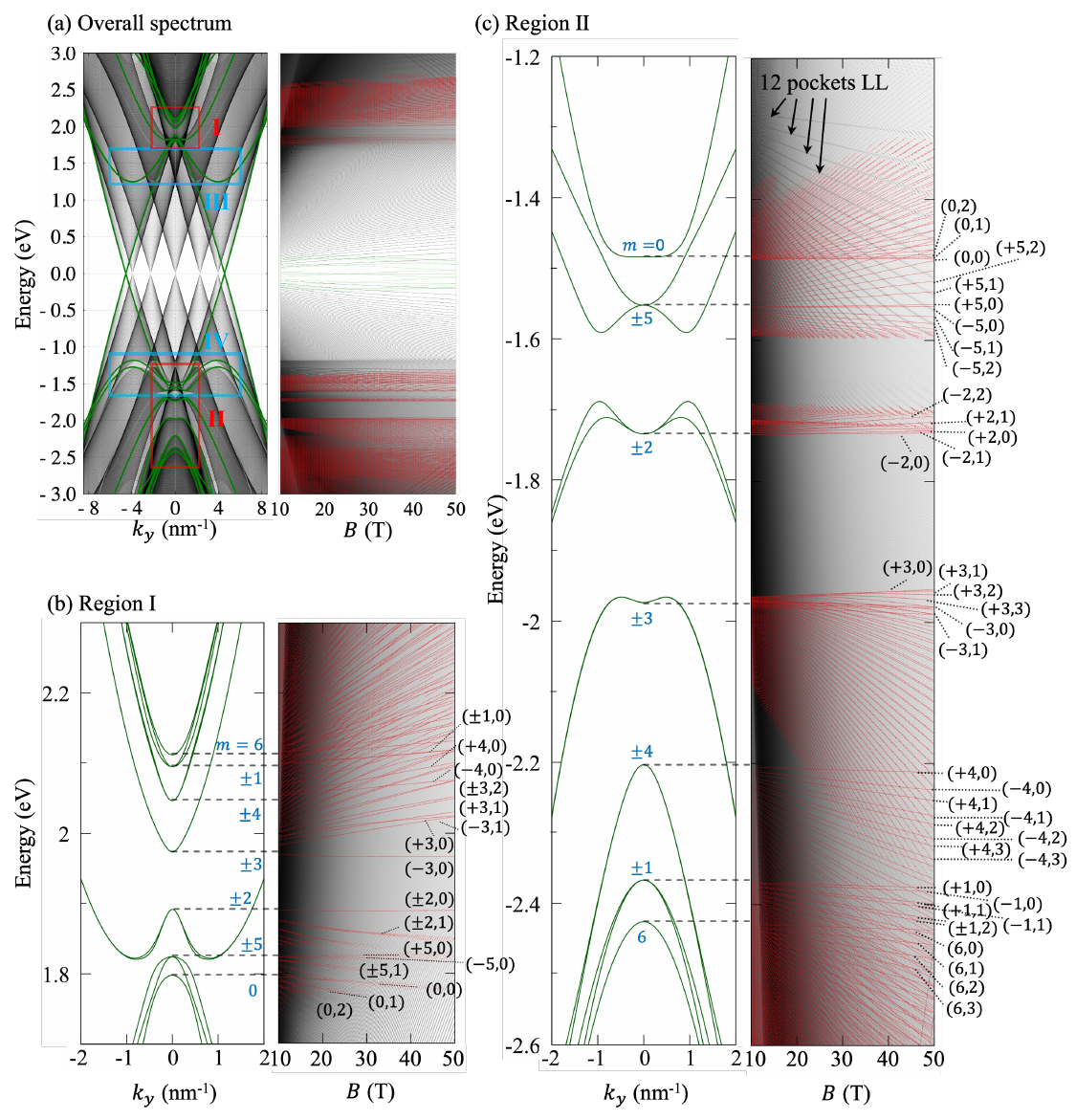}
    \caption{
    (a) Left: Quasi-band structure of 30$^\circ$ TBG projected onto $k_y$, with the green line indicating the band at $k_x = 0$.
    Right: Corresponding Landau levels in the same energy range. Red dashed lines mark the series of Landau levels associated with the band edges in the quasi-band structure (left panel). Green lines represent spurious levels due to the finite cutoff in the numerical calculation.
    (b) Enlarged view of Region I in the conduction band (red box in (a)). Each band is labeled by angular momentum $m$, and the Landau levels can be indexed as $(m, n)$, where $n$ denotes the Landau level index within each $m$ band.
    (c) Enlarged view of Region II in the valence band of (a), in the same manner as (b).
    }
    \label{fig_landau_all}
\end{figure*}

Figure~\ref{fig_landau_all} summarizes the Landau level structure of 
$30^\circ$ TBG calculated using the effective Hamiltonian. 
Figure~\ref{fig_landau_all}(a) presents the overall spectrum: 
the left panel shows the quasi-band structure at zero magnetic field, 
while the right panel displays the Landau level spectrum in the corresponding energy window, 
plotted as a function of magnetic field ranging from $10$~T to $50$~T. 
In the region $|E|\lesssim 1$~eV, we observe a series of Landau levels exhibiting the characteristic 
$\sqrt{B}$ dependence of monolayer graphene. 
This reflects the fact that the quasi-band structure of $30^\circ$ TBG 
retains the Dirac cone of the monolayer in the low-energy regime.
Green lines indicate spurious levels arising from the finite cutoff in the Landau level index, which are numerical artifacts.

In the higher-energy region $|E| > 1$~eV, we observe 
distinctive features reflecting the quasi-band structure of $30^\circ$ TBG. 
Figures~\ref{fig_landau_all}(b) and (c) show enlarged views of the energy regions~I, and II, respectively, 
as indicated in panel~(a). 
In these plots, the red lines denote a series of Landau levels associated with the band edges at $\bm{k} = 0$ 
in the quasi-band structure (shown in the left panel). 
We label these levels by $(m,n)$, 
where $m$ is the angular-momentum index of the corresponding quasi-band 
and $n$ is the Landau-level index. 
The identification proceeds as follows: 
each level has a well-defined total angular momentum $l = m + n$, 
since the spectrum is computed independently for each $l$. 
For a given level, $n$ is determined from the Landau index at which the wavefunction has the largest weight, 
and $m$ is then obtained as $m = l - n$. 
The red coloring is applied only when the wavefunction is well localized in a specific $n$, such that the squared weight on a single $n$ (summed over sublattices) exceeds~0.2.

We observe a clear correspondence between the quasi-band dispersions and the 
Landau-level structures. 
For example, the $m = \pm 5$ branches in Fig.~\ref{fig_landau_all}(c) 
exhibit positive and negative quadratic dispersions touching at the origin, 
which give rise to a series of electron-like and hole-like Landau levels centered at the same energy. 
The flat band bottom of the $m=0$ branch results in a dense cluster of low-lying Landau levels with weak magnetic-field dependence. 
Another notable feature is the splitting of Landau levels within the quasi-band pairs of angular momentum $\pm m$. 
For instance, the nearly degenerate parabolic bands with $m=\pm 4$ [Fig.~\ref{fig_landau_all}(d)] 
generate distinct series of Landau levels with a pronounced splitting. 
This splitting can be attributed to the Zeeman effect of the orbital magnetic moment.

\begin{figure}
    \centering
    \includegraphics[width = 1.\hsize]{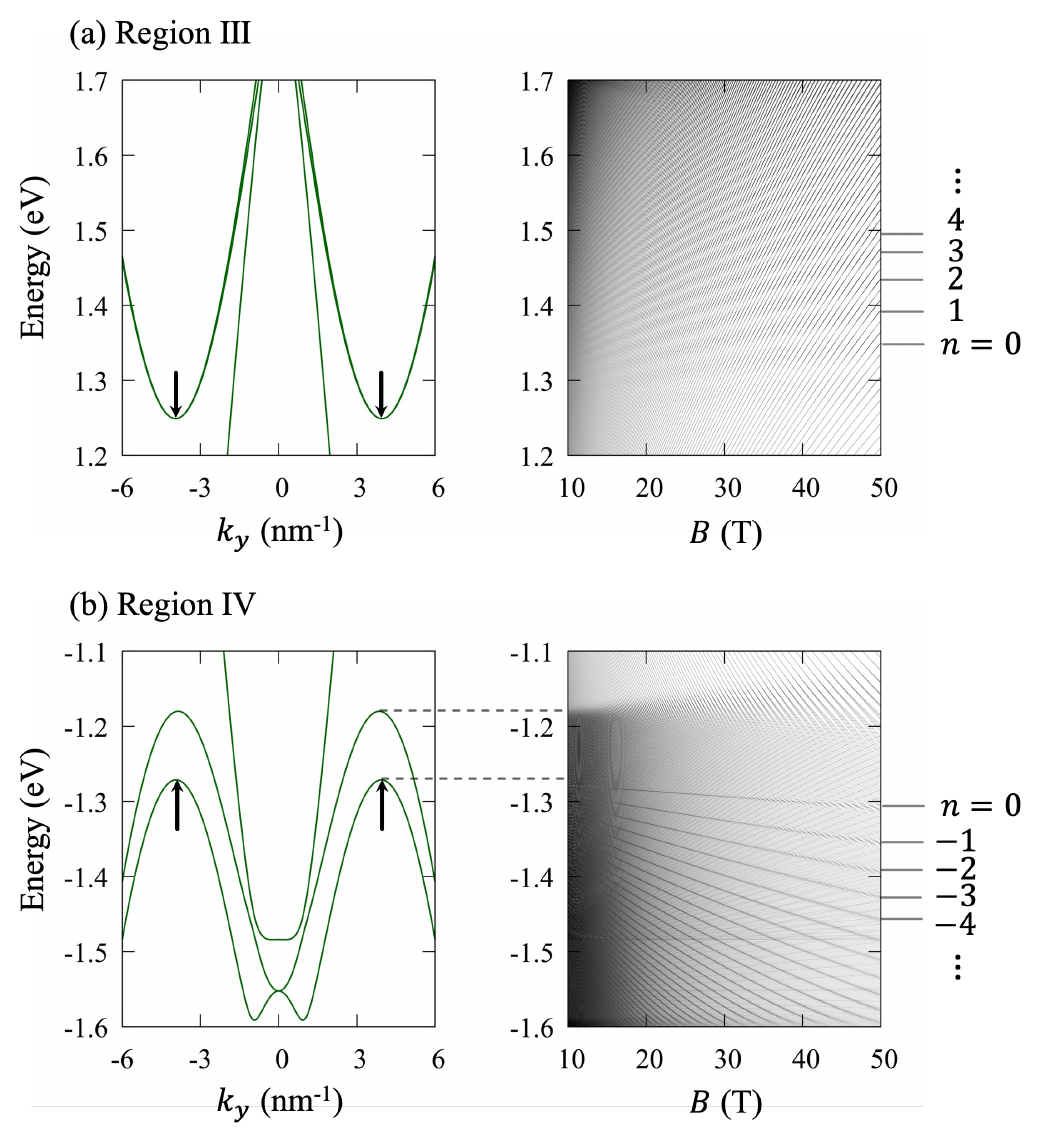}
    \caption{
    Quasi-band structure (left) and corresponding Landau levels (right) in Region III of Fig.~\ref{fig_landau_all}(a).
    (b) Same as (a), but for Region IV.
    }
    \label{fig_12-fold_pocket}
\end{figure}

On the other hand, we also find Landau levels that are not directly associated with the band structures near $\bm{k} = 0$. 
These appear as a sparse, downward-going fan indicated by black arrows in Fig.~\ref{fig_landau_all}(c). 
To clarify their origin, we show in Fig.~\ref{fig_12-fold_pocket}(b) the Landau-level spectrum together with the quasi-band structure over a wider momentum range, corresponding to region~IV in Fig.~\ref{fig_landau_all}(a). 
These levels correspond to the twelve off-center hole pockets arranged in a 12-fold symmetry, as shown in Fig.~\ref{fig_band_30tbg}(c). 
Consequently, the Landau levels are also 12-fold degenerate (per spin), in contrast to the nondegenerate levels associated with the pocket at $\bm{k} = 0$.  
A similar structure is also found in the electron part (region~III) as shown in Fig.~\ref{fig_12-fold_pocket}(a), although the Landau levels are significantly broadened due to the smaller energy gap in the band structure.
Notably, the level structure in region IV is consistent with previous results obtained using the approximant method \cite{yu2020electronic}.  
Importantly, the present approach further enables us to identify the microscopic origin of these levels by directly relating them to the quasi-band structure.

%%%%%%%%%%%%%%%%%%%%%%%%%%%%%%%%%
\section{Optical absorption}
\label{sec_magopt}

\subsection{Formulation}
\label{subsec_form_LLopt}

The dynamical conductivity of the $30^\circ$ TBG in the magnetic field
can be evaluated  by applying the Kubo formula to the Landau level eigenstates 
derived in the previous section. It reads,
\begin{align}
\mathrm{Re}  \sigma_{xx}(\omega)
   & = 
    %- \frac{i \hbar e^2}{V} N_B
    i \hbar e^2 N_B
    \sum_{n, n'}
    \sum_{l,l'}
    \sum_{X,X'}
    \nonumber\\
  &  
    \frac{f(\varepsilon_{n',l',X'}) - f(\varepsilon_{n,l,X})}
    {\varepsilon_{n',l',X'} - \varepsilon_{n,l,X}}
    \frac{
    \left|
    \bra{w^{X'}_{l',n'}} v_x \ket{w^X_{l,n}}
    \right|^2}
    {
    \hbar \omega + \varepsilon_{n',l',X'} - \varepsilon_{n,l,X}
    }
\end{align}
where $N_B = eB/\hbar$ is the number of states per area in a single Landau level.
Here the velocity operator $v_x$ is obtained by differentiating the Hamiltonian Eq.~\eqref{eq_H_ring} by the quasi-momentum as,
\begin{equation}
    v_x
        = \frac{1}{\hbar}
        \frac{\partial \mathcal{H}(\bm{k})}{\partial k_{x}},
\label{eq_vx}
\end{equation}
and applying the substitution Eq.~\eqref{eq_sbst}.
For the present system,
the matrix element of $v_x$ can be immediately obtained from Eq.~\eqref{eq_H_reduced_2nd} as
\begin{align}
    & \bra{w^{X'}_{l',n'}} v_x \ket{w^X_{l,n}} 
     = v \Big[
    (\tilde{V}_1\,\delta_{l',l-1}
   +  \tilde{V}_1^\dagger\,\delta_{l',l+1})\delta_{n',n} 
   \nonumber\\
    & \qquad
    +2\Delta_B \sqrt{n+1}\, \tilde{V}_2\,\delta_{l',l-1}\delta_{n',n+1} 
   +  2\Delta_B \sqrt{n}\, \tilde{V}_2^\dagger\,\delta_{l',l+1}\delta_{n',n-1} 
   \Big],
    \label{eq_v_LL_basis}
\end{align}
The velocity operator only has matrix elements between $l$ and $l \pm 1$ indicating that
the optical transition of the linearly polarized light only takes place between these states. This result aligns naturally with the conservation of angular momentum, noting that a linearly polarized photon is composed of the angular momentum $\pm 1$ (in units of $\hbar$).

The validity of applying Eq.~\eqref{eq_vx} to quasi-periodic systems is not obvious, and it can be verified as follows.  
For the original tight-binding Hamiltonian $H$, the velocity operator is defined as
$v_\mu = (1/i\hbar)[x_\mu,H]$.  
Using the interlayer matrix element of the Hamiltonian in Eq.~\eqref{eq_interlayer_H},  
the corresponding matrix element of $v_\mu$ is evaluated as
\begin{align}
& \langle \bm{k}^{(2)},X',2 |v_\mu| \bm{k}^{(1)},X,1 \rangle 
\nonumber\\
&=
 \frac{1}{N}
\sum_{\bm{R} \in \bm{R}^{(1)}_{X}} 
\sum_{\bm{R}' \in \bm{R}^{(2)}_{X'}} 
 e^{i\bm{k}^{(1)}\cdot\bm{R}-i\bm{k}^{(2)}\cdot\bm{R}'}
\langle\bm{R}'|
\frac{1}{i\hbar}(x_\mu H-H x_\mu)
|\bm{R}\rangle
\nonumber\\
&=
\frac{1}{N}
\sum_{\bm{R} \in \bm{R}^{(1)}_{X}} 
\sum_{\bm{R}' \in \bm{R}^{(2)}_{X'}} 
e^{i\bm{k}^{(1)}\cdot\bm{R}-i\bm{k}^{(2)}\cdot\bm{R}'}
 \frac{-1}{i\hbar}(R'_\mu -R_\mu)
 T(\bm{R}'-\bm{R})
%\langle\bm{R}'|H|\bm{R}\rangle
\nonumber\\
&= -\frac{1}{\hbar}\sum_{\bm{G} \in \bm{G}^{(1)}} \sum_{\bm{G}' \in \bm{G}^{(2)}} 
t_\mu(\bm{k}^{(1)}+\bm{G}^{(1)})
\, e^{-i\bm{G}^{(1)}\cdot\bm{\tau}^{(1)}_{X}
	+i\bm{G}^{(2)}\cdot\bm{\tau}^{(2)}_{X'}}
    \nonumber\\
    &\hspace{4cm}\times \delta_{ \bm{k}^{(1)}+\bm{G}^{(1)},\,\bm{k}^{(2)}+\bm{G}^{(2)}},
\label{eq_interlayer_v}
 \end{align}
where $t_\mu(\bm{q}) = \partial t(\bm{q})/\partial q_\mu$.  
In the third line of Eq.~\eqref{eq_interlayer_v}, we used the identity
\[
-i r_\mu T(\bm{r}+d_0 \bm{e}_z) = \frac{S_0}{(2\pi)^2} \int d^2\bm{q}\, t_\mu(\bm{q}) e^{i\bm{q}\cdot\bm{r}},
\]
which can be readily derived by differentiating Eq.~\eqref{eq_hop_tq} with $q_\mu$.
By comparing Eqs.~\eqref{eq_interlayer_H} and \eqref{eq_interlayer_v},
and considering that
$\bm{k}^{(1)} = \bm{k} + \bm{G}^{(2)}$ with the quasi-momentum
$\bm{k}$ [Eq.~\eqref{eq_wavespace}],
we obtain
\begin{align}
    \langle \bm{k}^{(2)},X',2 |v_\mu| \bm{k}^{(1)},X,1 \rangle 
    =   \frac{\partial}{\partial k_\mu} 
    \langle \bm{k}^{(2)},X',2 |H| \bm{k}^{(1)},X,1 \rangle,
\end{align}
which leads to Eq.~\eqref{eq_vx}.

\begin{figure*}
    \centering
    \includegraphics[width =0.9\hsize]{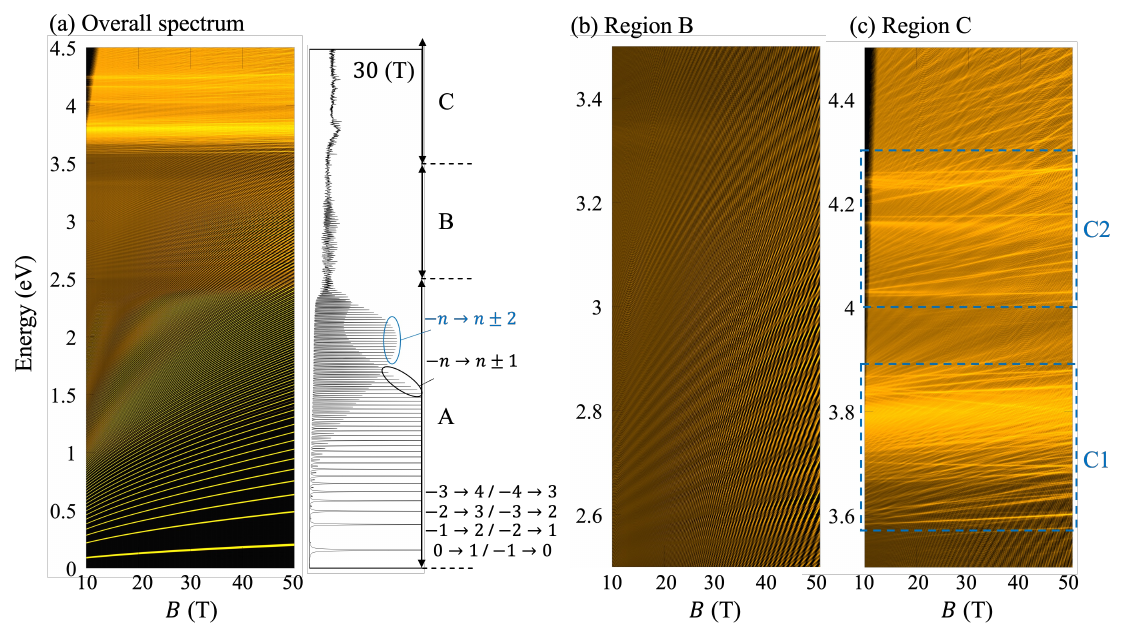}
    \caption{ 
(a) Density plot of the optical absorption spectrum ${\rm Re},\sigma_{xx}$ of 30$^\circ$ TBG as a function of magnetic field $B$ and transition energy $\hbar\omega$. The right panel shows a line cut at $B=30$ T. We identify three regions, A, B, and C, based on qualitative differences in their spectral features. Region A exhibits a structure reminiscent of monolayer graphene, while regions B and C display more complex patterns arising from the characteristic 12-fold quasi-band structure.
(b), (c) Enlarged views focusing on regions B and C, respectively.
}
    \label{fig_opt_all}
\end{figure*}

\begin{figure*}
    \centering
    \includegraphics[width =1.\hsize]{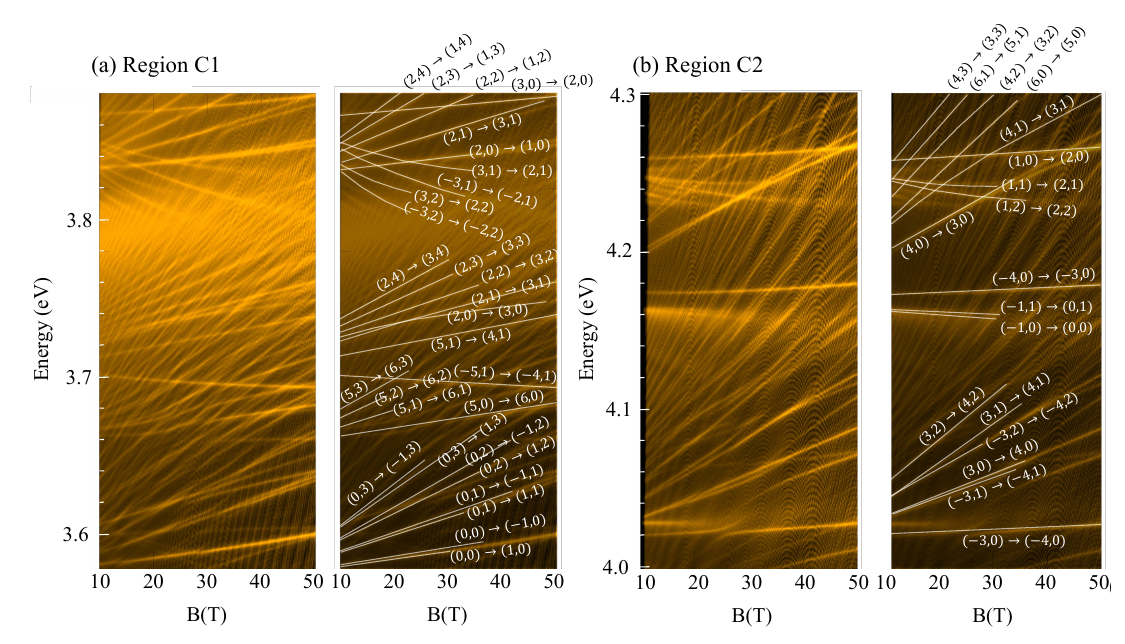}
    \caption{
    Enlarged views of the optical absorption spectrum in Fig.~\ref{fig_opt_all}(c), focusing on Regions C1 and C2.
The right panel in each figure shows the transition indices $(m, n) \rightarrow (m', n')$, with $(m, n)$ and $(m', n')$ labeling the initial and final Landau levels in the valence and conduction bands.
    }
    \label{fig_opt_part}
\end{figure*}

\subsection{Optical absorption Spectrum}
\label{subsec_sp_LLopt}

In Fig.~\ref{fig_opt_all}(a), we show the density plot of the dynamical conductivity ${\rm Re}\,\sigma_{xx}(\omega, B)$ 
in the magnetic field range of $10~{\rm T} <B < 50~{\rm T}$ and 
the frequency range $0 <\hbar\omega < 4.5$~eV,
with the spectrum at $B =30~{\rm T}$.
We divide the energy range into three regions, region A, B, and C,
based on the qualitative differences in the spectral structure.

In region A, we observe a structure reminiscent of the optical spectrum of monolayer graphene, aligning with the band structure that preserves the monolayer's Dirac cone.
The monolayer's Landau level is labeled by $n=0,\pm1,\pm2,\cdots$, where the negative and positive branches correspond to the conduction and valence Dirac cone. \cite{mcclure1956Diamagnetism,gusynin2005Unconventional}
Here the main absorption peaks in the region A are associated with the transition from $-n$ to $n\pm 1$, which are the transitions in the ideal Dirac Hamiltonian with a perfect circular symmetry.
We also observe a series of $-n$ to $n\pm 2$, which grows significantly in the energy range $E > \sim 1$ eV.
These transitions 
(more generally, $n \to n\pm1+3m$) are allowed when the circular symmetry of the Dirac cone is reduced to three-fold rotational symmetry, which is mainly contributed by the interlayer coupling in the present system.

Regions B and C exhibit a complex structure distinct from monolayer's spectrum, as seen in the detailed plot in Fig.~\ref{fig_opt_part}(a) and (b).
The complex features arise from the characteristic Landau level structure of the
30$^\circ$ TBG, which was discussed in the previous section.
The absorption peaks in the region C
can be labeled as $(m,n)\to (m',n')$,
which are the initial and final states in the excitation.
Figure \ref{fig_opt_part} presents the magnified
plot for the region C with identified labels.
We find that most of the relevant peaks yield to the selection rule,
\begin{equation}
    (m, n) \rightarrow (m \pm 1, n).
    \label{eq_mn_selection_rule}
\end{equation}
These transitions correspond to the optical excitation 
in the absence of the magnetic field,
which occurs from the band $m$ to $m\pm 1$.
The selection rule Eq.~\eqref{eq_mn_selection_rule}
obviously satisfies the conservation of the total angular momentum $l \to l\pm 1$, where $l=m+n$.
More generally, the transitions are also feasible from $(m, n)$ to $(m + k \pm 1, n - k)$,
while these additional peaks are found to be notably smaller than the main series of Eq.~\eqref{eq_mn_selection_rule}.
Lastly, in region B,
a spectrum structure reflecting the 12-fold Landau levels, as seen in Fig.~\ref{fig_12-fold_pocket}(b), emerges.
The radial spectrum structure shown in Fig.~\ref{fig_opt_all}(b) corresponds to 
transitions between the 12-fold Landau levels,
from the valence band (region IV) 
to the conduction band (region III).

\section{Conclusion}
\label{sec_concl}

We have developed a compact theoretical framework to compute Landau levels in quasi-periodic stacks of 2D materials, and applied it to the dodecagonal quasicrystal realized by $30^\circ$ twisted bilayer graphene. 
Starting from the quasi-band description, the coupling to a magnetic field is implemented by the minimal substitution for the quasi-momentum as in conventional periodic systems, 
thereby enabling bulk Landau-level spectra at modest computational cost. 
This framework provides a transparent physical interpretation of the Landau levels in terms of the quasi-band structure, allowing them to be understood as quantized orbits of quasi-band pockets.  
This perspective enables a direct identification of the microscopic origin of characteristic spectral features, such as dense stacks of nearly field-independent Landau levels associated with flat quasi-band edges, as well as 12-fold degenerate levels arising from off-center pockets.  
The Landau levels are labeled by a pair of quantum numbers $(m,n)$, where $m$ denotes the 12-fold quasi-band angular momentum, $n$ the Landau-level index, and the total angular momentum $l=m+n$ is conserved.

We further computed the magneto-optical conductivity and established selection rules governed by the 12-fold angular momentum. 
The dominant absorption lines follow the rule $(m,n) \rightarrow (m\pm 1,n)$ [Eq.~\eqref{eq_mn_selection_rule}], i.e., 
$\Delta l=\pm1$ carried by a linearly polarized photon.
%while weaker satellites originating from higher-order couplings appear at $(m,n) \rightarrow (m+k\pm1,n-k)$. 
In the frequency regime associated to the quasicrystalline band structure, the spectrum fragments into bundles that track the $(m,n)$ quantum numbers and the 12-pocket fans. 
On the experiment side, 
characteristic features constitute robust spectroscopic fingerprints that can be sought in high-field infrared/THz probes.

Our approach can be applied to other quasiperiodic van der Waals heterostructures for which the quasi-band description is available. 
It opens a route to bulk quantum magneto-optics in nonperiodic solids without resorting to real-space supercells. 
In summary, we provide a gauge-invariant Landau-level theory for quasiperiodic stacks and demonstrate it for $30^\circ$ TBG. 
The quasiband-based interpretation, together with the angular-momentum selection rules in magneto-optics, establishes a coherent framework for understanding and designing Landau-level structures and magneto-optical responses in quasicrystalline moiré materials.

\bibliography{reference}

\end{document}